\documentclass{elsart}
\usepackage{epsf}
\usepackage{amssymb}

\newcommand\Ref[1]     {Ref.\,\cite{#1}}
\newcommand\Refs[1]    {Refs.\,\cite{#1}}
\newcommand\Eqn[1]     {Eq.\,(\ref{#1})}
\newcommand\Eqns[2]    {Eqs.\,(\ref{#1}) and~(\ref{#2})}

\newcommand\Figs[2]    {Figs.\,\ref{#1} and \ref{#2}}

\newcommand\nn         {\nonumber}

\def\beq{\begin{equation}}
\def\eeq{\end{equation}}
\def\beeq{\begin{eqnarray}}
\def\eeeq{\end{eqnarray}}
\def\la{\langle}
\def\ra{\rangle}
\newcommand\bom[1]     {{\mbox{\boldmath $#1$}}}

\newcommand\as         {\ensuremath{\alpha_{\mathrm{s}}}}

\newcommand\msbar      {\ensuremath{{\overline {\rm MS}}}}
\newcommand\muR[1]     {\ensuremath{\mu_R^{#1}}}
\newcommand\muF[1]     {\ensuremath{\mu_F^{#1}}}

\newcommand\rB         {\mathrm{B}}
\newcommand\rR         {\mathrm{R}}
\newcommand\rV         {\mathrm{V}}
\newcommand\rC         {\mathrm{C}}
\newcommand\rA         {\mathrm{A}}
\newcommand\rLO        {\mathrm{LO}}
\newcommand\rNLO       {\mathrm{NLO}}

\newcommand{\eps}      {\varepsilon}        

\newcommand{\rd}       {{\mathrm{d}}}

\newcommand\kT         {k_\perp}
\newcommand\pdf{parton distribution function}
\newcommand\LO{LO}
\newcommand\NLO{NLO}

\newcommand\xB    {\ensuremath{x_{\rm B}}}
\newcommand\ycut  {\ensuremath{y_{\rm cut}}}

\newcommand\Kout  {\ensuremath{K_{\rm out}}}

\newcommand\Qhs   {\ensuremath{Q_{\rm H.S.}}}

\begin{document}

\begin{frontmatter}
\begin{flushright}
hep-ph/0511328 \\
\end{flushright}

\title{Three-jet event-shapes in lepton-proton scattering at
next-to-leading order accuracy}
\author[UniZh]{Zolt\'an Nagy} and
\author[DE]{Zolt\'an Tr\'ocs\'anyi}
\address[UniZh]{University of Z\"urich,
Winterthurerstrasse 190,
CH-857 Z\"urich, Switzerland}
\address[DE]{University of Debrecen
and Institute of Nuclear Research of the Hungarian Academy of Sciences,
H-4001 Debrecen P.O.Box 51, Hungary}
\date{\today}

\begin{abstract}
We compute the differential distributions of two three-jet event-shape
observables at the next-to-leading order accuracy at fixed values of
the DIS kinematic variables. The observable $\Kout$ measures the
out-of-event-plane momentum. The other observable $y_3$ is the maximum
value of the $\ycut$ resolution variable for which an event is
classified as three-jet event. We also show the dependence of the
fixed-order predictions on the renormalization and factorization
scales. The radiative corrections are in general large and depend
on the value of the DIS kinematic range.
\end{abstract}

\end{frontmatter}

\section{Introduction}

The analysis of event-shape observables in $e^+e^-$-annihilation and
in deeply-inelastic lepton-proton scattering (DIS) proved to be a
powerful method to study Quantum Chromodynamics (QCD) \cite{DSevshape}.
The standard QCD analysis of event-shapes consists of matching the
next-to-leading order (NLO) and resummed next-to-leading logarithmic
(NLL) predictions, possibly improved with analytic predictions for the
power corrections (PC). Such matched predictions also describe the
distributions of multi-jet rates in $e^+e^-$ annihilation with high
accuracy \cite{NTQCD98}. Thus, it is interesting to investigate whether
similar level of accuracy can also be achieved in predicting
distributions of multi-jet observables in DIS.

Two-jet\footnote{In counting the number of jets in this paper, we neglect
the ubiquitous beam-jet.} event-shape observables in DIS have been
thoroughly analyzed both theoretically \cite{DSDIS} and experimentally
\cite{HERAevshapes}. Three-jet rates have been also analyzed
\cite{HERA3jets} based on computations valid at the NLO accuracy
\cite{NTdisPRL}. Resummed prediction to the three-jet rates have not yet
been computed, therefore matched predictions are not available for jet
rates. However, much progress has been achieved in computing resummed
predictions at the NLL accuracy for three-jet event shapes
\cite{Kout,caesar,caesarpage} that are sensitive to large angle soft
emission and thus exhibit rich geometry-dependent structure.  These
developments in theory encouraged the experimenters to consider
three-jet event shapes \cite{Everett}.  However, predictions to
three-jet event shapes at the NLO accuracy have not been computed
yet. In this letter we aim to fill this gap.

It is well-known that the validity of fixed-order predictions is rather
constrained. In the perturbative expansion of the distribution of an
observable $O$, logarithmic terms $\alpha_s^m \log^n O$ appear which
require all-order resummation if the value of $O$ is small. In analyzing
the data, most of the statistics lie in the range of small $O$, thus
the computation of resummed predictions is indispensable for the  
experimental analysis. The observables we choose to compute are
those for which resummed predictions are known.

The first three-jet event-shape observable in DIS that has been
computed at the NLL accuracy is the $\Kout$ variable, that measures the
out-of-event-plane QCD radiation.  The observable $\Kout$ was defined
in \Ref{Kout} as the sum of the momentum components perpendicular to
the event plane,
\beq
\Kout = \sum_h
|p_h^{\rm out}|\:.
\label{Kout}
\eeq
The summation extends over all particles (hadrons in the experiment,
partons in the theoretical computation).
The event plane is spanned by the proton three-momentum $\vec{p}$ and
the unit vector $\vec{n}$ that defines the thrust major axis in the
plane perpendicular to the beam,
\beq
T_M = \max_{\vec{n}} \frac1Q \sum_h
|\vec{p}_h\cdot \vec{n}|\:,
\qquad
\vec{n}\cdot \vec{p} = 0\,.
\eeq

Recently the {\sc caesar} program has been published
\cite{caesar,caesarpage} that can be used for computing cross sections
of two- and three-jet event shapes%
\footnote{It can also be used for computing dijet event shapes in hadronic
collisions.}
in an automatic way. In particular, the distribution of
the $y_3$ observable, that is defined to be the largest value of the jet
resolution variable $\ycut$ such that the event is clustered into three
jets, is also known to NLL accuracy \cite{caesarpage}. For defining the jets,
the computation uses the $\kT$-clustering algorithm of \Ref{kTclusDIS}.

In this paper we compute the distributions of the two three-jet event
shape observables in DIS for which resummed predictions already exist,
but the fixed-order radiative corrections have not been computed
before.  We use the {\sc nlojet++} program \cite{nlojet}.

\section{Computation of fixed-order predictions}

The next-to-leading order cross section for electron-proton
scattering into three jets is the convolution of the parton density
function of the incoming proton and the hard scattering cross section, 
\beq
\label{had-xsec}
\nn
\sigma(p, q) = \sum_{a}
\int_0^1\!\!\rd\eta \, f_{a/P}(\eta,\mu_F^2)
\,\left[\sigma^{\rLO}_{a}(\eta p, q) 
  + \sigma^{\rNLO}_{a}(\eta p, q)\right]\:,
\eeq
where $p^\mu$ and $q^\mu$ are the four-momenta of the incoming proton and
the exchanged virtual photon, respectively, $f_{a/P}(\eta,\mu_F^2)$ is
the density of the parton of type $a$ in the incoming proton at
momentum fraction $\eta$ and factorization scale $\mu_F$. The
corresponding parton level cross sections are
\beq
\sigma^{\rLO}_{a}(p, q) \equiv \int_3 d\sigma^{\rB}_{a}(p, q) 
= \int_3 \rd\Gamma^{(3)}\la|M^{(3)}_{a}|^2\ra J^{(3)}\:,
\label{par-xsec-lo}
\eeq
and the next-to-leading order correction is sum of three terms
\beeq
\sigma^{\rNLO}_{a}(p, q) &&\equiv \int \rd\sigma^{\rNLO}_{a}(p, q)
\nn\\&&
= \int_4\!\rd\sigma^{\rR}_{a}(p, q) + \int_3\!\rd\sigma^{\rV}_{a}(p, q)
 + \int_3\!\rd\sigma^{\rC}_{a}(p, q)\:,\qquad
\label{par-xsec-nlo}
\eeeq
where $\rd\sigma^{\rR}$ and $\rd\sigma^{\rV}$ are the real and virtual
contributions to the partonic cross section. The contribution
$\rd\sigma^{\rC}$ represents the collinear-subtraction counter term. The
pole structure of this term is unique, while its finite part depends on
the factorization scheme. We use the \msbar\ scheme as defined
precisely in \Ref{CSdipole}. The parton density functions are also
scheme dependent, so that the scheme-dependence cancels in the hadronic
cross section of \Eqn{had-xsec}.

The three integrals on the right hand side of \Eqn{par-xsec-nlo} are 
separately divergent but their sum is finite provided the jet function
$J^{(m)}$ defines a collinear and infrared safe quantity, which
formally means that
\beq
J^{(4)} \longrightarrow J^{(3)}\;\;,
\eeq
whenever the four-parton and three-parton configurations are
kinematically degenerate (regions of one unresolved parton). In
addition, to define a three-jet observable, both $J^{(4)}$ and
$J^{(3)}$ have to vanish if less than three partons are resolved.
The presence of the singularities means that the separate pieces have
to be regularized and the divergences have to be cancelled. We use
dimensional regularization in $d=4-2\eps$ dimensions in which case the
divergences are replaced by double and single poles of the form
$1/\eps^2$ and $1/\eps$. We assume that ultraviolet renormalization of
all Green functions to one-loop order has been carried out, so the
divergences are of infrared origin. In order to get the finite sum one
has to rearrange the various contributions by subtracting and adding
the same, in $d=4$ dimensions singular terms to the three contributions
in \Eqn{par-xsec-nlo} so that each becomes separately finite in $d=4$
dimensions.
 
The essence of this rearrangement is to define a single subtraction
term $\rd\sigma^{\rA}$ that regularizes the divergences in the real
term which comes form the unresolved soft and collinear regions. Thus,
the three singular integrals in Eq. (\ref{par-xsec-nlo}) are substituted
by three finite ones:
\beq
\sigma^{\rNLO}_{a}(p, q) = \sigma^{\rNLO\{4\}}_{a}(p, q)
+ \sigma^{\rNLO\{3\}}_{a}(p, q)
+ \int_0^1 \rd x\,\hat{\sigma}^{\rNLO\{3\}}_{a}(x, xp, q)\:,
\label{final-xsec}
\eeq
where the four-parton integral is given by
\beq
\sigma^{\rNLO\{4\}}_{a}(p, q) = \int_4
\left[
  \rd\sigma^{\rR}_{a}(p, q)_{\eps=0}
- \rd\sigma^{\rA}_{a}(p, q)_{\eps=0}
\right]\:.
\label{sigma-NLO4}
\eeq
We have two three-parton contributions to the NLO correction. The
second term on the right hand side of Eq. (\ref{final-xsec}) is the sum
of the one-loop contribution and a Born term convoluted with a universal
singular factor $\bom{I}$,
\beq
\sigma^{\rNLO\{3\}}_{a}(p, q) = \int_3
\left[\rd\sigma^{\rV}_{a}(p, q)
+ \rd\sigma^{\rB}_{a}(p, q)\otimes \bom{I}\right]_{\eps=0}\:.
\eeq
The factor $\bom{I}$ contains all the $\epsilon$ poles which come from the 
$\rd\sigma^{\rA}$ and  $\rd\sigma^{\rC}$ terms that are necessary to
cancel the (equal and with opposite sign) poles in $\rd\sigma^{\rV}$. The
$\otimes$ operation means correlations in colour space. The last term
in Eq. (\ref{final-xsec}) is a finite remainder, in the form of
a convolution, that is left after factorization of initial-state
collinear singularities into the non-perturbative parton distribution
functions, 
\beeq
&&
\int_0^1 \rd x \, \hat{\sigma}^{\rNLO\{3\}}_{a}(x, xp, q) =
\nn \\ && \qquad
 \sum_{a'}\int_0^1 \rd x
\int_3
\left[\rd\sigma^{\rB}_{a'}(xp, q)\otimes [\bom{P}(x) + \bom{K}(x)]^{aa'}
\right]_{\eps=0}\:,
\eeeq
where the $x$-dependent functions $\bom{P}$ and $\bom{K}$ are similar
(but finite for $\eps\to 0$) to the factor $\bom{I}$. These functions
are universal, that is, they are independent of the scattering process
and of the jet observables.  

There are many ways to define the $d\sigma^{\rA}$ subtraction
term, but all must lead to the same finite next-to-leading order
correction.  In computing the \NLO\ corrections to multijet cross
sections the dipole subtraction scheme of Catani and Seymour
\cite{CSdipole} is a convenient formalism. It is used both in the
{\sc disent} and the {\sc nlojet++} programs.  The subtraction scheme
applied in the {\sc nlojet++} program is modified slightly as compared
to the original one in \cite{CSdipole} in order to have a better
control on the numerical computation. The main idea is to cut the phase
space of the dipole subtraction terms as introduced in
\Ref{NT4jet}.  We thus define the $\rd\sigma^\rA$ local
counter term as
\beeq
&&
\rd\sigma_{a}^\rA = \sum_{\{4\}} \rd\Gamma^{(4)}(p_a,q,p_1,...,p_4)
  \frac1{S_{\{4\}}}
\nn\\ &&\qquad
\times\Bigg\{\sum_{\mathrm{pairs}\atop i,j} \sum_{k\neq i,j}
  D_{ij,k}(p_a,q,p_1,\dots,p_4)
  J^{(3)}(p_a,\dots,\tilde{p}_{ij},\tilde{p}_{k},\dots)
  \Theta(y_{ij,k} < \alpha)
\nn\\ &&\qquad\qquad
+ \sum_{\mathrm{pairs}\atop i,j}
  D_{ij}^a(p_a,q,p_1,\dots,p_4)
  J^{(3)}(\tilde{p}_a,\dots,\tilde{p}_{ij},\dots) 
  \Theta(1-x_{ij,a} < \alpha)
\nn\\ &&\qquad\qquad
+ \sum_{i\neq k}
  D_k^{ai}(p_a,q,p_1,\dots,p_4)
  J^{(3)}(\tilde{p}_a,\dots,\tilde{p}_k,\dots) 
  \Theta(u_{i} < \alpha)
\Bigg\}\:,
\label{dipole-terms}
\eeeq
where $\rd\Gamma^{(4)}$ is the four-parton phase space including all the
factors that are QCD independent, $\sum_{\{4\}}$ denotes the sum over
all configurations with $4$ partons and $S_{\{4\}}$ is the Bose symmetry
factor of the identical partons in the final state. The $D_{ij,k}$,
$D_{ij}^a$ and $D_k^{ai}$ functions are the dipole factors
given in \Ref{CSdipole}. The function $J^{(3)}$ is the jet function
which acts over the three-parton dipole phase space. The variables
$y_{ij,k}$, $x_{ij,a}$ and $u_{i}$ are the dipole variables used for
defining the exact factorization of the phase space \cite{CSdipole}.
The parameter $\alpha \in (0,1]$ controls the volume of the dipole phase
space. The case of $\alpha = 1$ means the full dipole subtraction. We can
speed up the computer program by choosing $\alpha \simeq 0.1$, which
keeps the subtraction in the vicinity of the singular regions, but avoids
the CPU-intensive computations of the dipole terms where those are not
necessary. Furthermore, checking that the predictions are independent
of the parameter $\alpha$, that sets the volume of the cut dipole phase
space, gives a strong control that indeed the same quantity is
subtracted from the real correction as added to the virtual one. 
Choosing $\alpha < 1$, the insertion operators $\bom{I}(\alpha, \eps)$,
$\bom{P}(\alpha)$, $\bom{K}(\alpha)$  depend on $\alpha$.  The explicit
expressions can be found in \Ref{Npp3jet}.  

The integrand of the NLO contribution $\sigma^{\rNLO\{4\}}_{a}$ in
\Eqn{sigma-NLO4} contains integrable square root singularities.
Integrating these singularities by simple Monte Carlo integration
technique (choosing random values of the integration variables
uniformly) is not efficient because the variance of the estimate of the
integral is formally infinite, therefore, the estimate of the
statistical error of the integral is unreliable.  To improve the
convergence of the Monte Carlo integral, in the {\sc nlojet++} program
the phase space is generated by multi-channel weighted phase space
generator \cite{Kleiss:1994qy}.

Once the phase space integrations are carried out, we write the NLO
jet cross section in the following form:
\beq
\sigma^{(J)} =
\sum_{a} \int_0^1 \rd\eta
\,f_{a/P}(\eta, \mu_F^2)\,\sigma_{a,\rNLO}^{(J)}
\left(p_a, q,\as(\mu_R^2), \muR 2/Q_{HS}^2, \muF 2/Q_{HS}^2\right)\:,
\label{signjet}
\eeq
where $\sigma_{a,\rNLO}^{(J)}$ represents the sum of the LO and NLO
contributions to the partonic cross section, given in
\Eqns{par-xsec-lo}{final-xsec} respectively, with jet function $J$.
In addition to the parton momenta and possible parameters of the jet
function, it also depends explicitly on the renormalized strong
coupling $\as(\muR 2)$, the renormalization and factorization scales
$\muR{} = x_R \Qhs$ and $\muF{} = x_F \Qhs$, where $\Qhs$ is the hard
scale that characterizes the parton scattering. The scale $\Qhs$ is
usually set event by event to a measurable energy scale of the event.
Furthermore, the cross section also depends on the electromagnetic
coupling, for which we used $\msbar$ running
$\alpha_{\scriptscriptstyle \rm EM}(Q^2)$ at the scale of the virtual
photon momentum squared, $Q^2 = - q^2$. 

The publicly available version of the {\sc nlojet++} program
\cite{nlojet} is based on the tree-level and one-loop matrix elements
given in \Refs{NT4jet,BDK}, crossed into the photon-parton channel. It
uses a C/C++ implementation of the LHAPDF library \cite{Giele:2001mr}
with CTEQ6 \cite{Pumplin:2002vw} \pdf s and with the corresponding $\as$
expression for the renormalized coupling which is included in this
library.  The CTEQ6 set was fitted using the two-loop running coupling
with $\as(M_{Z^0}) = 0.118$.

\section{Results}

\begin{figure}
\centerline{
\epsfxsize=8cm \epsfbox{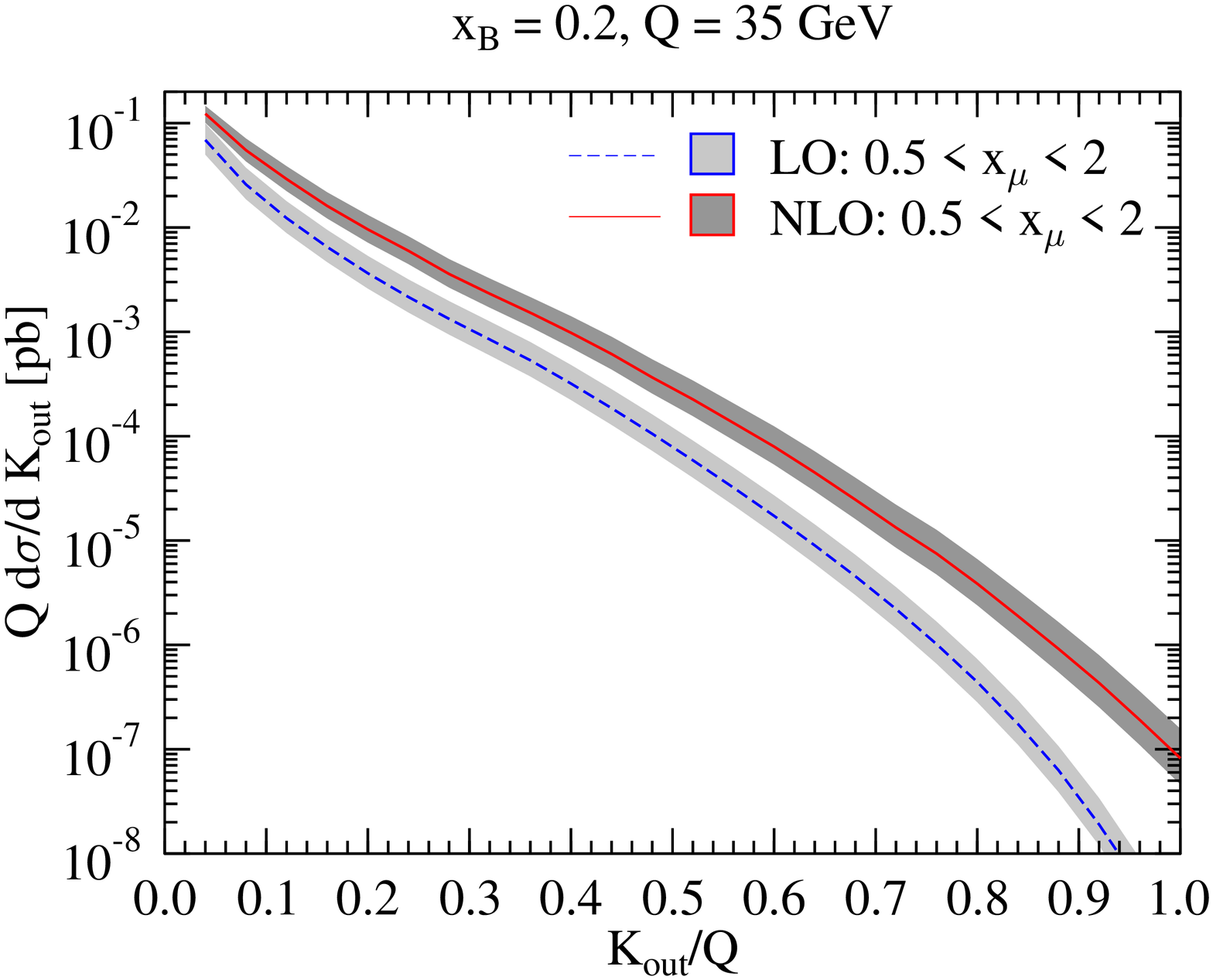}
\epsfxsize=8cm \epsfbox{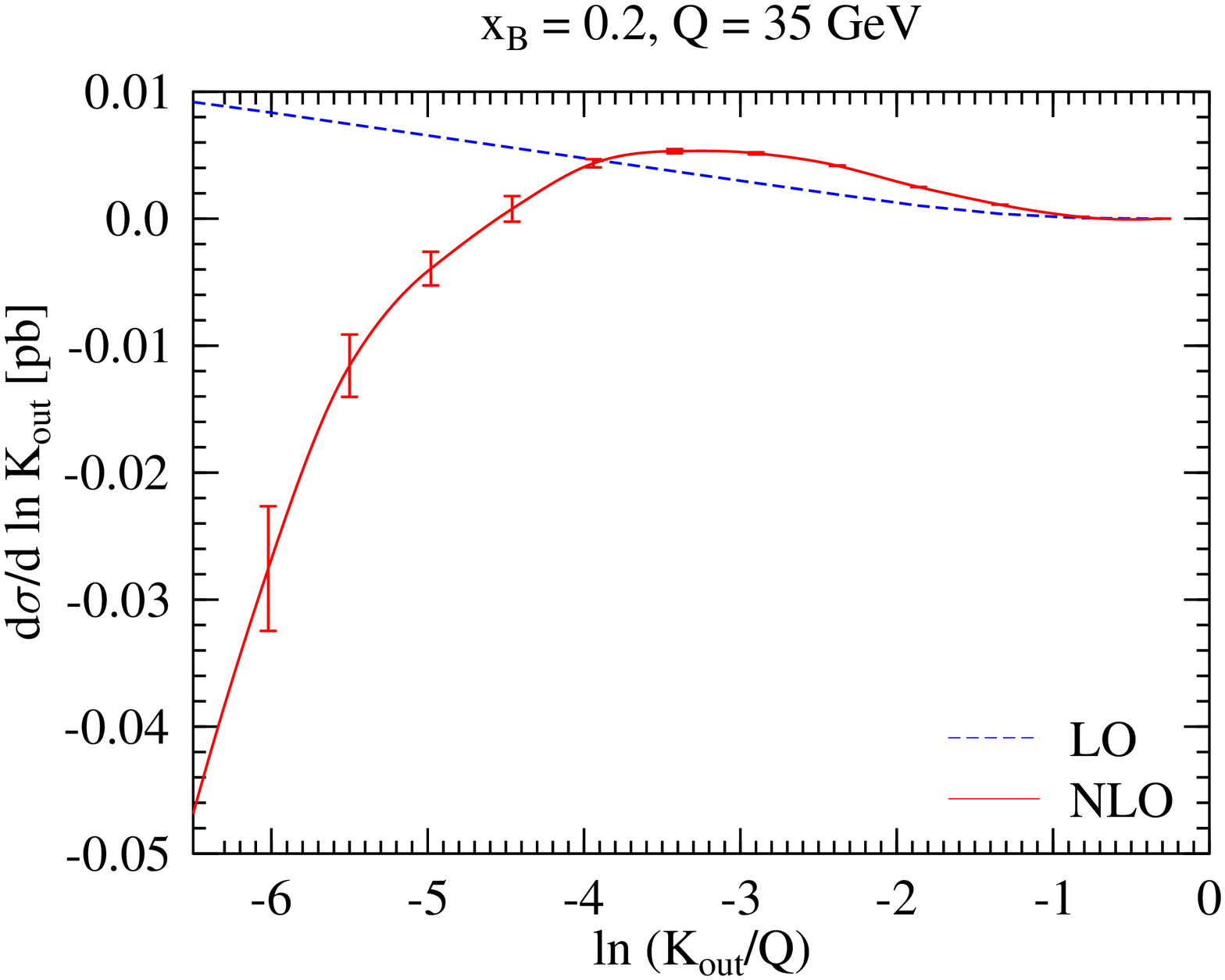}
}
\centerline{
\epsfxsize=8cm \epsfbox{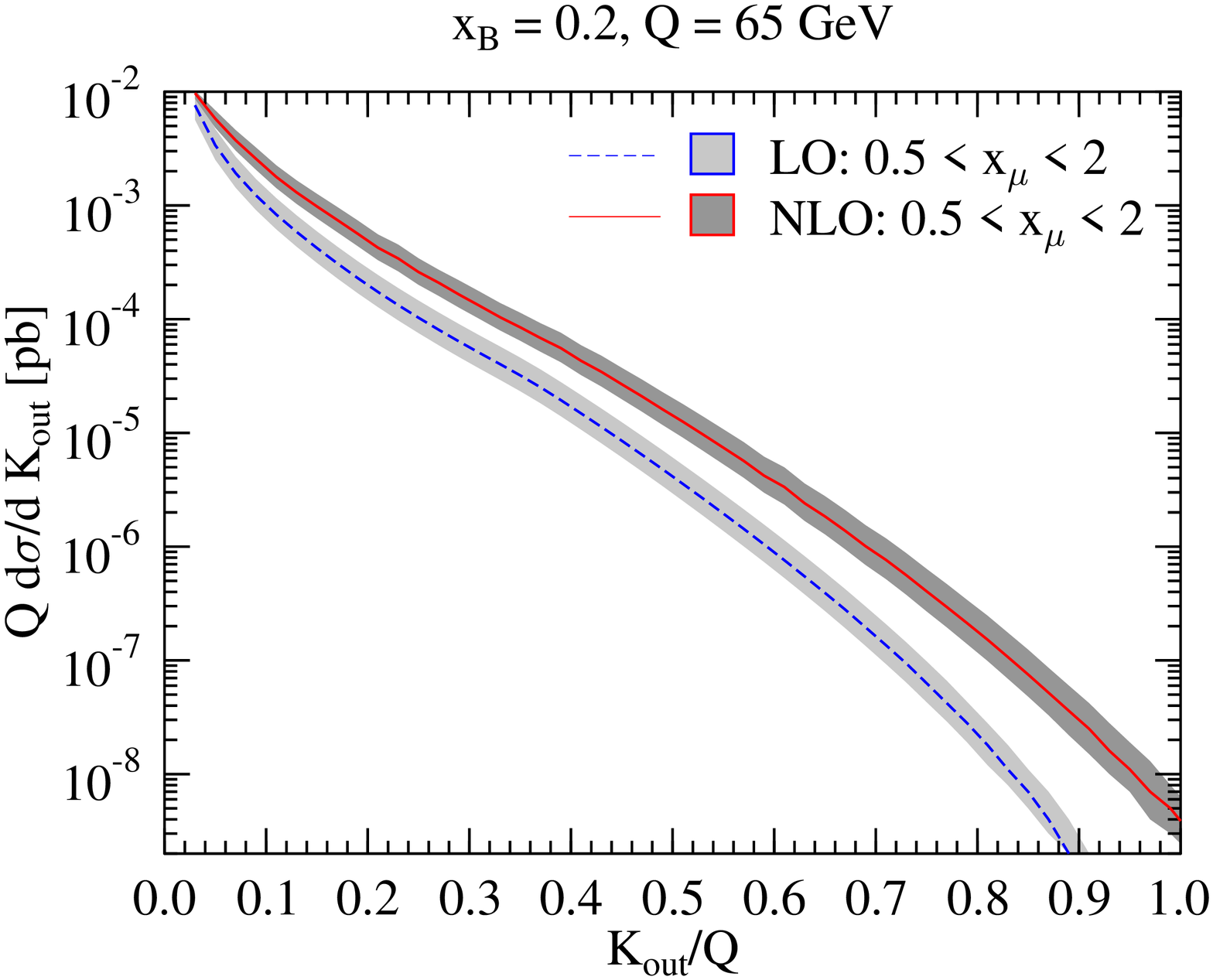}
\epsfxsize=8cm \epsfbox{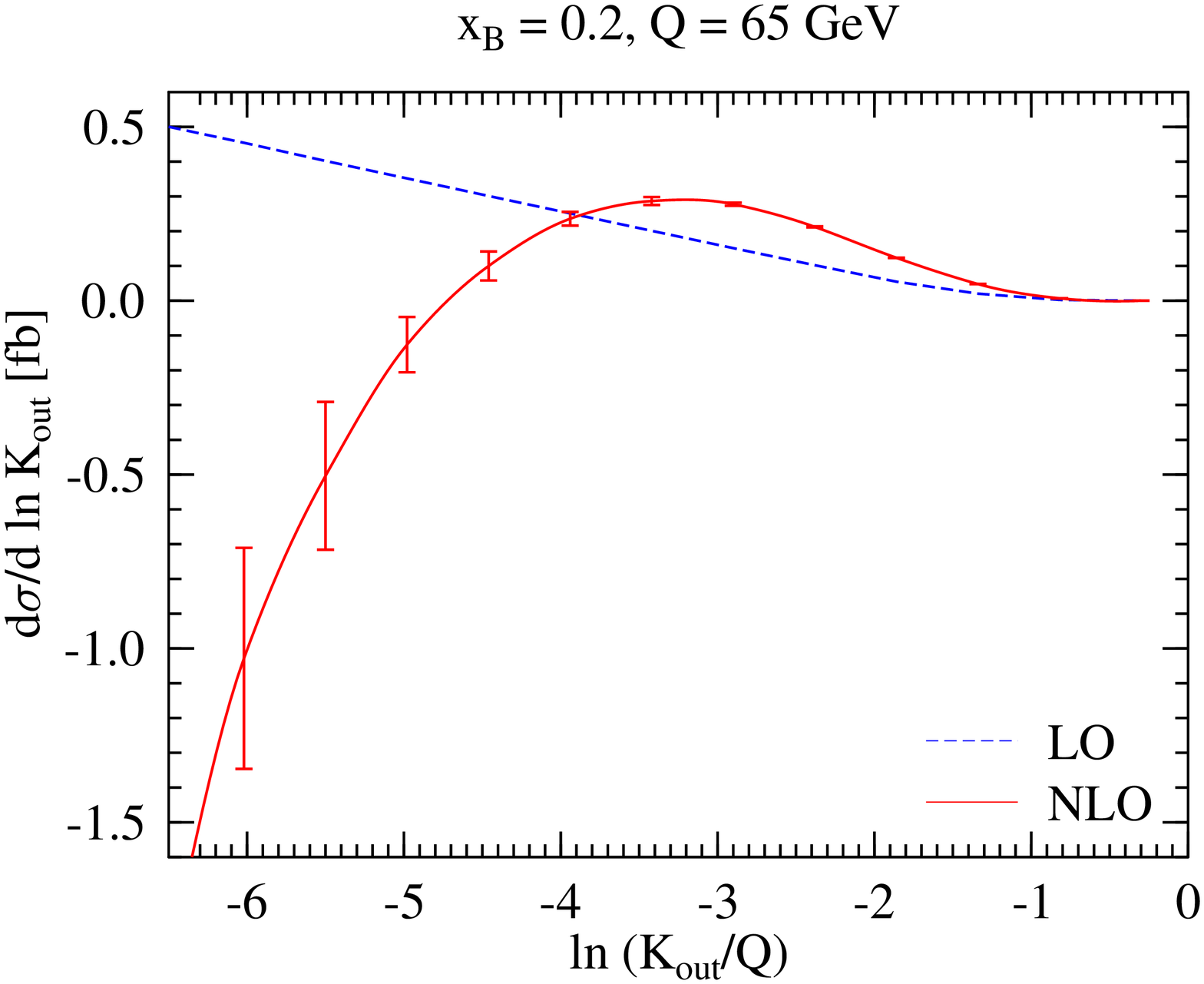}
}
\centerline{
\epsfxsize=8cm \epsfbox{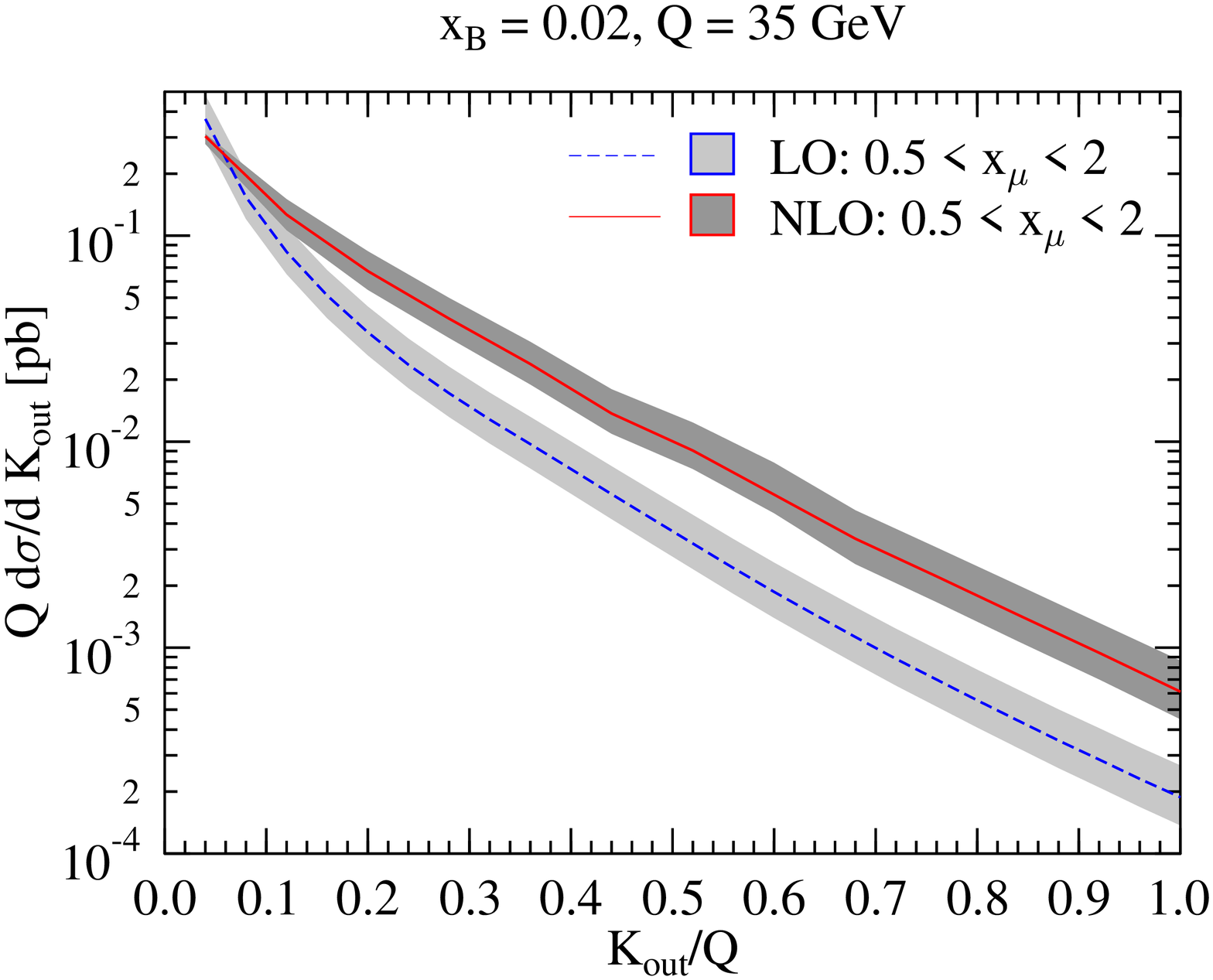}
\epsfxsize=8cm \epsfbox{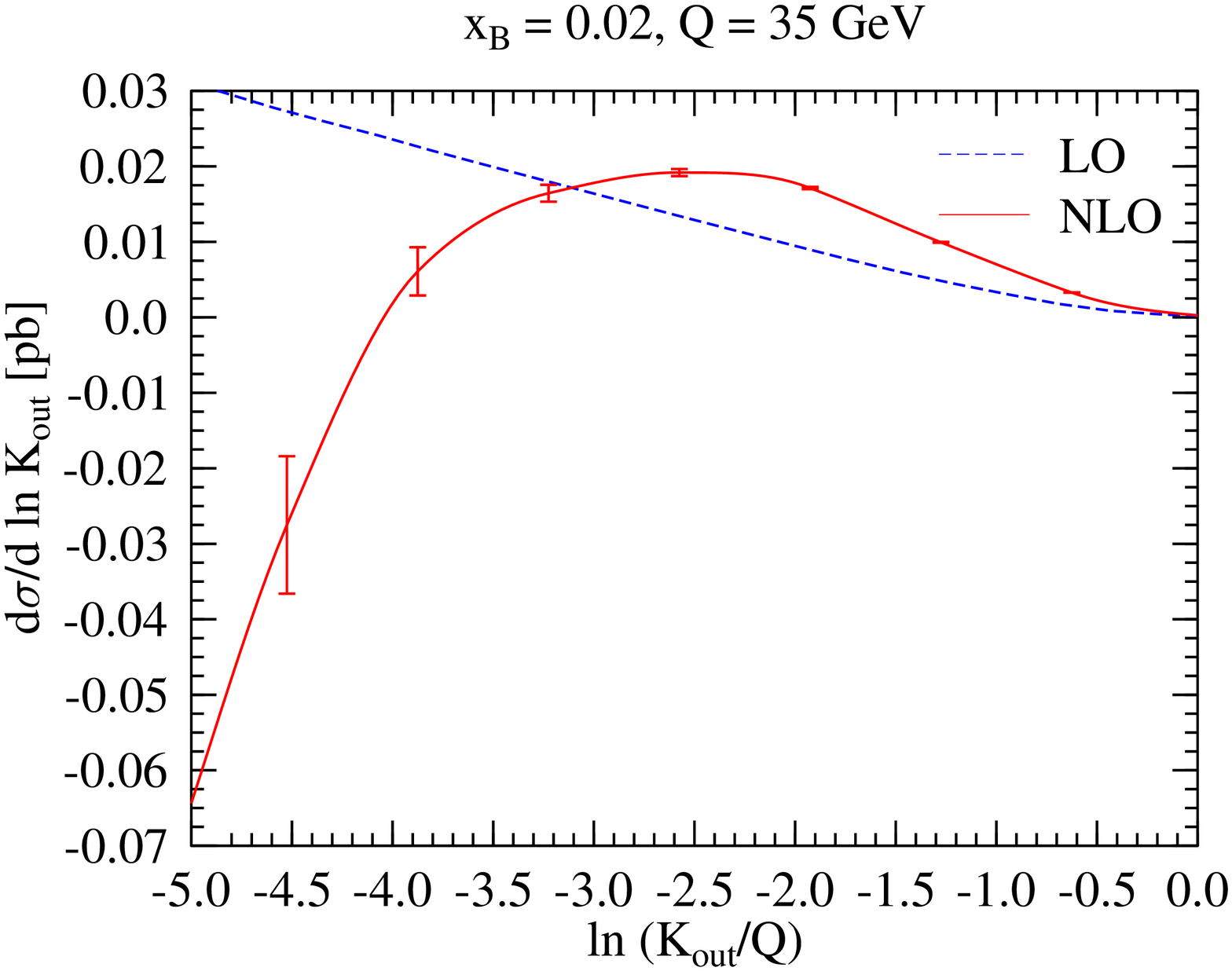}
}
\caption{The differential distribution of the $\Kout$ observable at
three different fixed values of the DIS kinematic variables. The left
panel shows the distributions as a function of $\Kout/Q$, the right
panel shows the distributions as a function of $\ln \Kout/Q$ in order to
exhibit the logarithmic dominance for small values of the observable.
The \LO\ predictions are shown with dashed lines, the \NLO\ predictions
are shown with solid lines.  The errorbars indicate the uncertainty of
the numerical integration that is negligible for the computation at LO
accuracy.}
\label{fig:Kout}
\end{figure}

We computed the distributions for fixed values of the DIS kinematic
variables $Q^2$ and $\xB$, as done in the resummation computations
\cite{caesarpage}.  We used three sets of values: $Q^2 = (35\,{\rm GeV})^2$
with $\xB=0.02$ and 0.2 and $Q^2 = (65\,{\rm GeV})^2$ with $\xB=0.2$. The
$Q^2 = (65\,{\rm GeV})^2$ with $\xB=0.02$ kinematical point is not
accessible at HERA.  In order to select events with at least two jets
with high transverse momenta ($p_\perp \sim Q$), we required $y_2 >
0.1$, where $y_2$ is the maximum value of the jet resolution variable
$\ycut$ such that the event can be classified as a two-jet one
\cite{Kout}. We employed a cut on the rapidity of the final-state
momenta in the Breit frame such that
\beq
\eta_i =
\frac12 \ln \left(1 + \frac{q\cdot p_i}{\xB\,p\cdot p_i}\right) < 3\,.
\eeq
\Figs{fig:Kout}{fig:y3} show the differential distributions for
$\Kout$ and $y_3$ observables.  The shaded bands correspond to the
range of scales $1/2 \le x_\mu \le 2$,
where $x_\mu^2 = \muR{2}/\Qhs^2 = \muF{2}/\Qhs^2$, with hard
scattering scale chosen to be $\Qhs^2 = Q^2$. We observe several
features of these plots.

Let us consider first the distributions in $\Kout$. We find that
the radiative corrections are in general large, thus the scale-dependence
reduces only relatively to the cross sections. The corrections also
depend strongly on the values of the DIS kinematic variables: they
increase with decreasing $Q^2$ and with decreasing $\xB$ (as seen from
the plots in the right panel).
They also increase with increasing value of $\Kout$ because the phase
space for events with large out-of-plane radiation with three partons in
the final state (at LO) is much smaller than that with four partons in
the final state (real corrections). The boundary of the phase space in
$\Kout$ depends on the value of $\xB$, increases with decreasing $\xB$,
and is about 20\,\% larger for the NLO computation than at LO. The
cross sections decrease rapidly with increasing $\Kout$. The rate of
this decrease also depends on $\xB$, being much quicker for larger
values of $\xB$ due to the smaller phase space. Nevertheless, the small
cross section for medium or large values of $\Kout$ leaves the small
$\Kout$-region for experimental analysis.  

In the small $\Kout$-region, the logarithmic contributions of the type
$\ln \Kout/Q$ are dominant as can be seen on the plots in the right
panel. At LO, the cross section behaves as $-\as(Q) \ln \Kout/Q$,
while at NLO the asymptotically leading term is $\as^2(Q) \ln^3 \Kout/Q$
for small values of $\Kout$, therefore, the fixed-order predictions diverge
with $\Kout \to 0$ with alternating signs, which makes the resummation
of these large logarithmic contributions mandatory.  Reliable
theoretical predictions can be obtained by matching the cross sections
valid at the NLO and NLL accuracy as described in \Ref{matching,DSDIS}.

Similar qualitative conclusions can be drawn from the $y_3$ distributions
with some important differences. Although the corrections are also large,
they are much smaller than in the case of the $\Kout$. Thus the reduction
in the scale-dependence can clearly be seen. An important reason for the
smaller corrections is that the phase space in $y_3$ is the same at LO as
at NLO. The size of the corrections depends more on the value of the
momentum transfer, less on $\xB$. However, the phase space in $y_3$ is
more dependent on $\xB$ (increases with decreasing $\xB$), than on $Q^2$
(decreases slightly with increasing $Q^2$). The cross section is sizable
again for small values of the event shape and the need for resummation in
this region is clearly seen on the right panel.

\begin{figure}
\centerline{
\epsfxsize=8cm \epsfbox{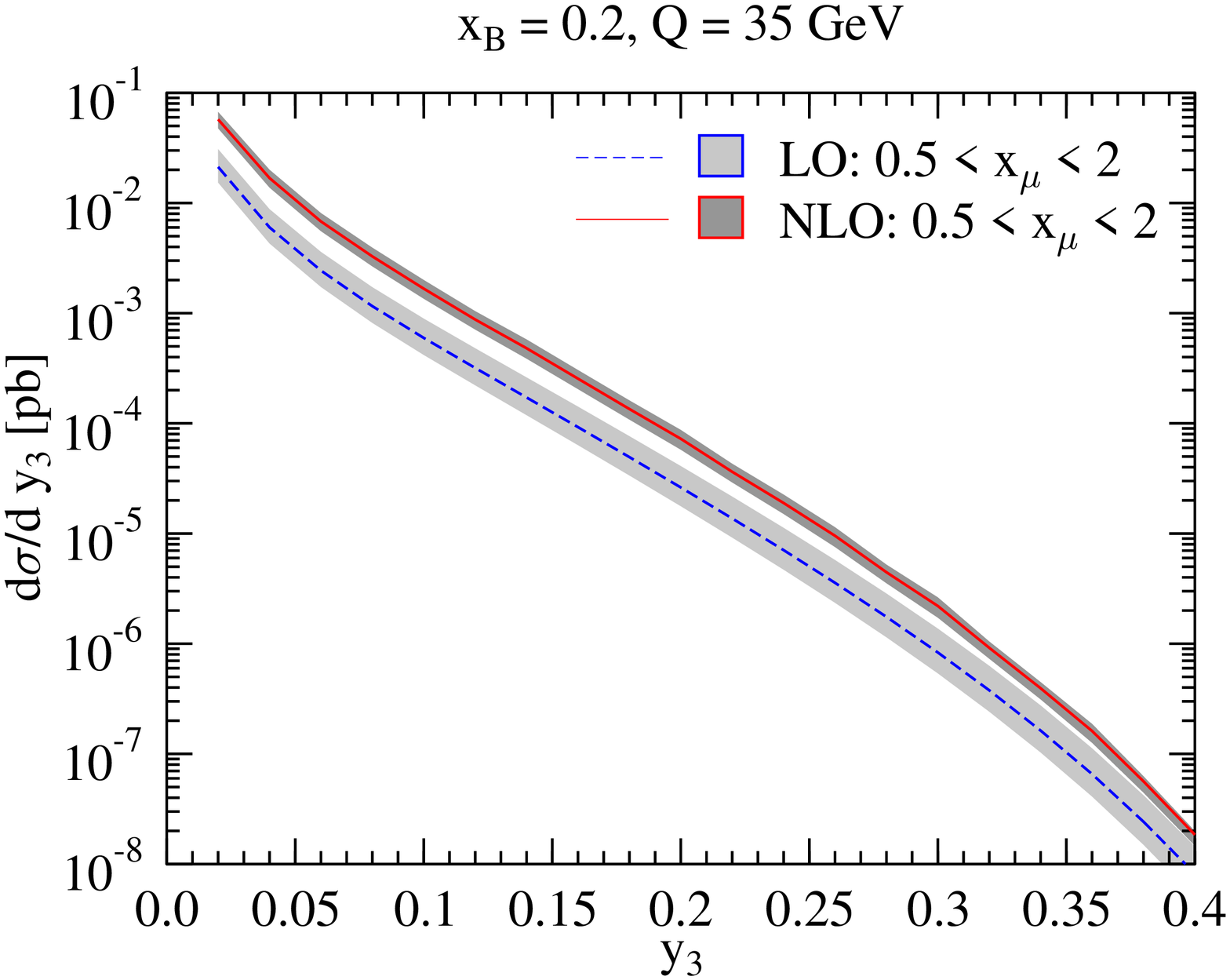}
\epsfxsize=8cm \epsfbox{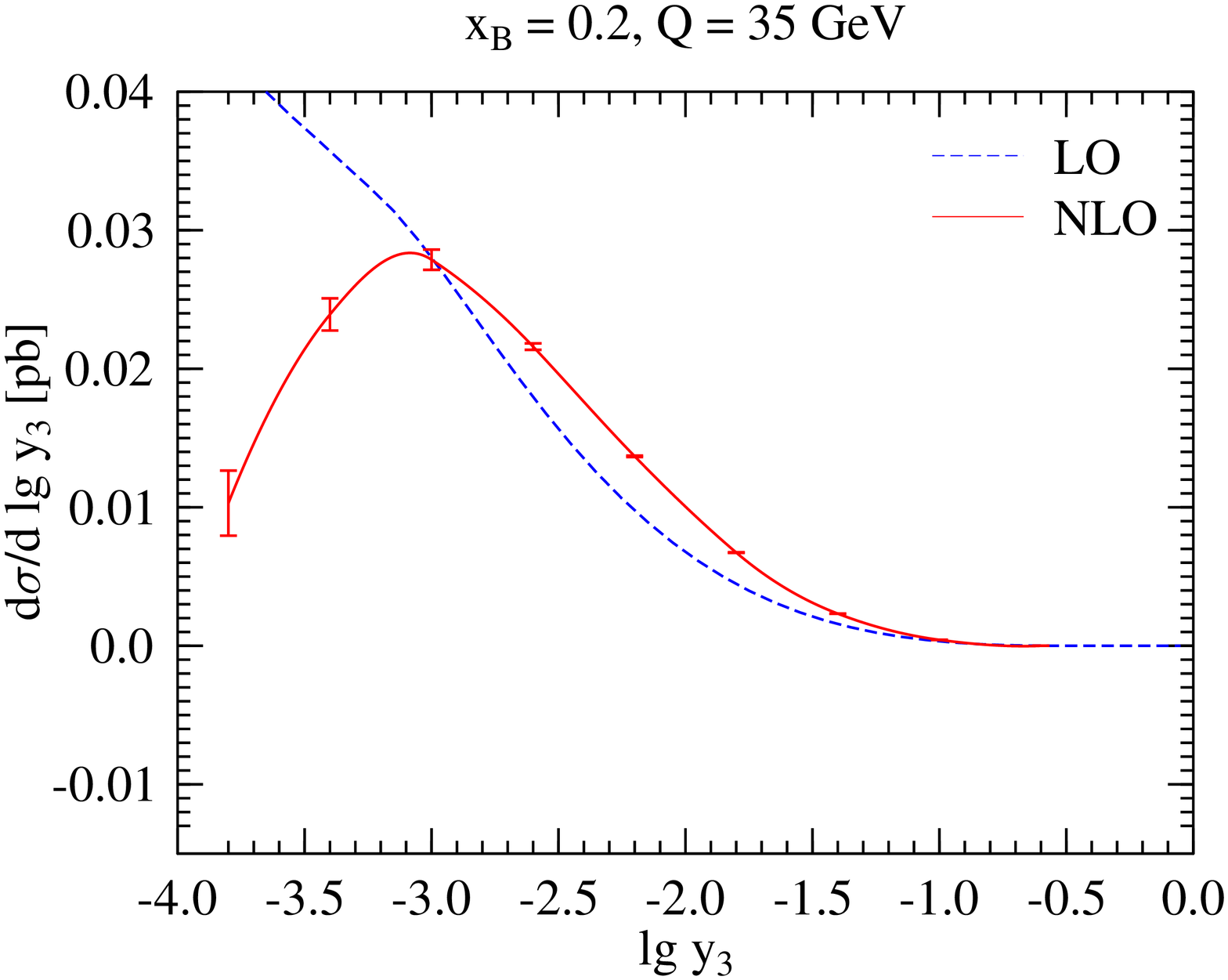}
}
\centerline{
\epsfxsize=8cm \epsfbox{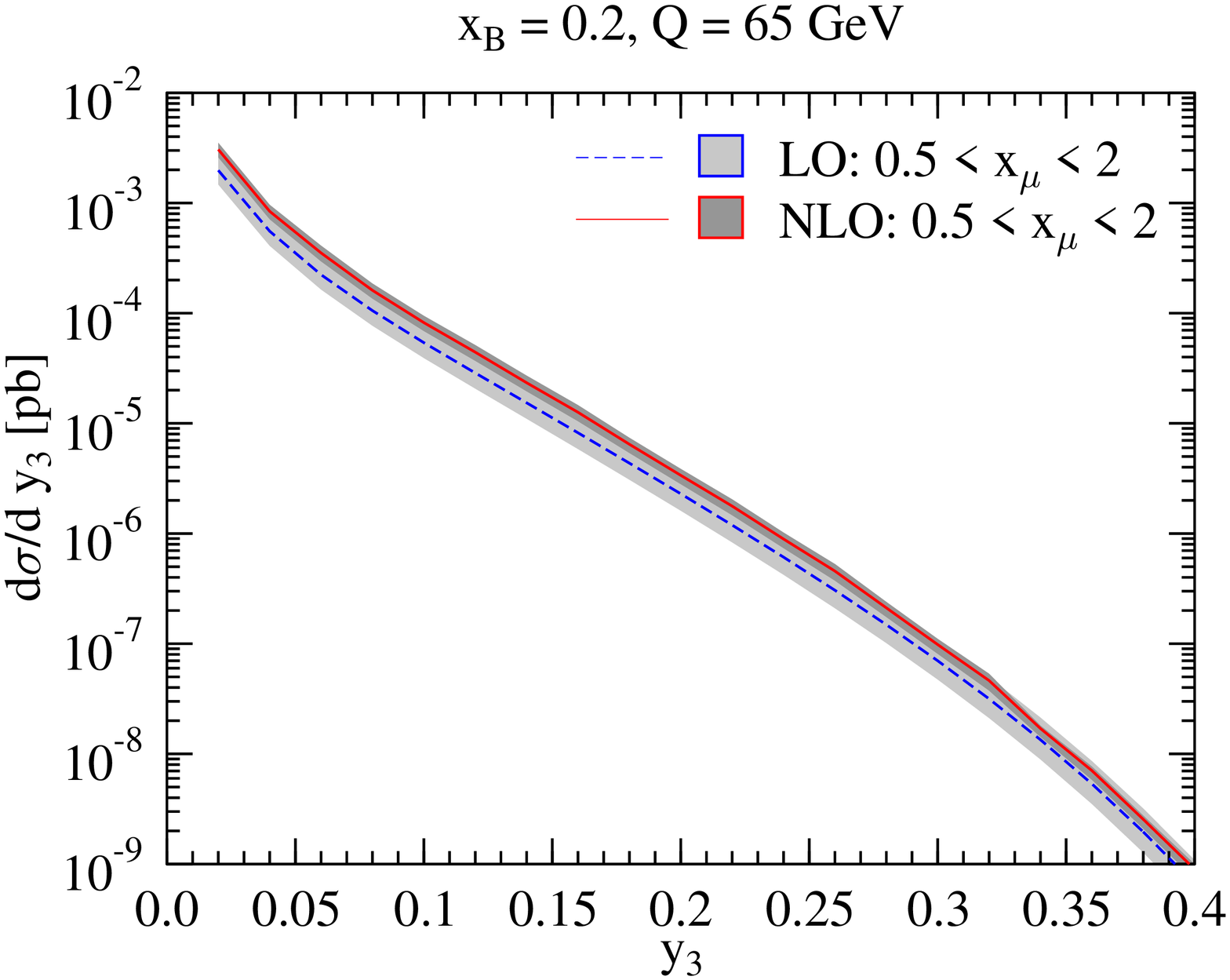}
\epsfxsize=8cm \epsfbox{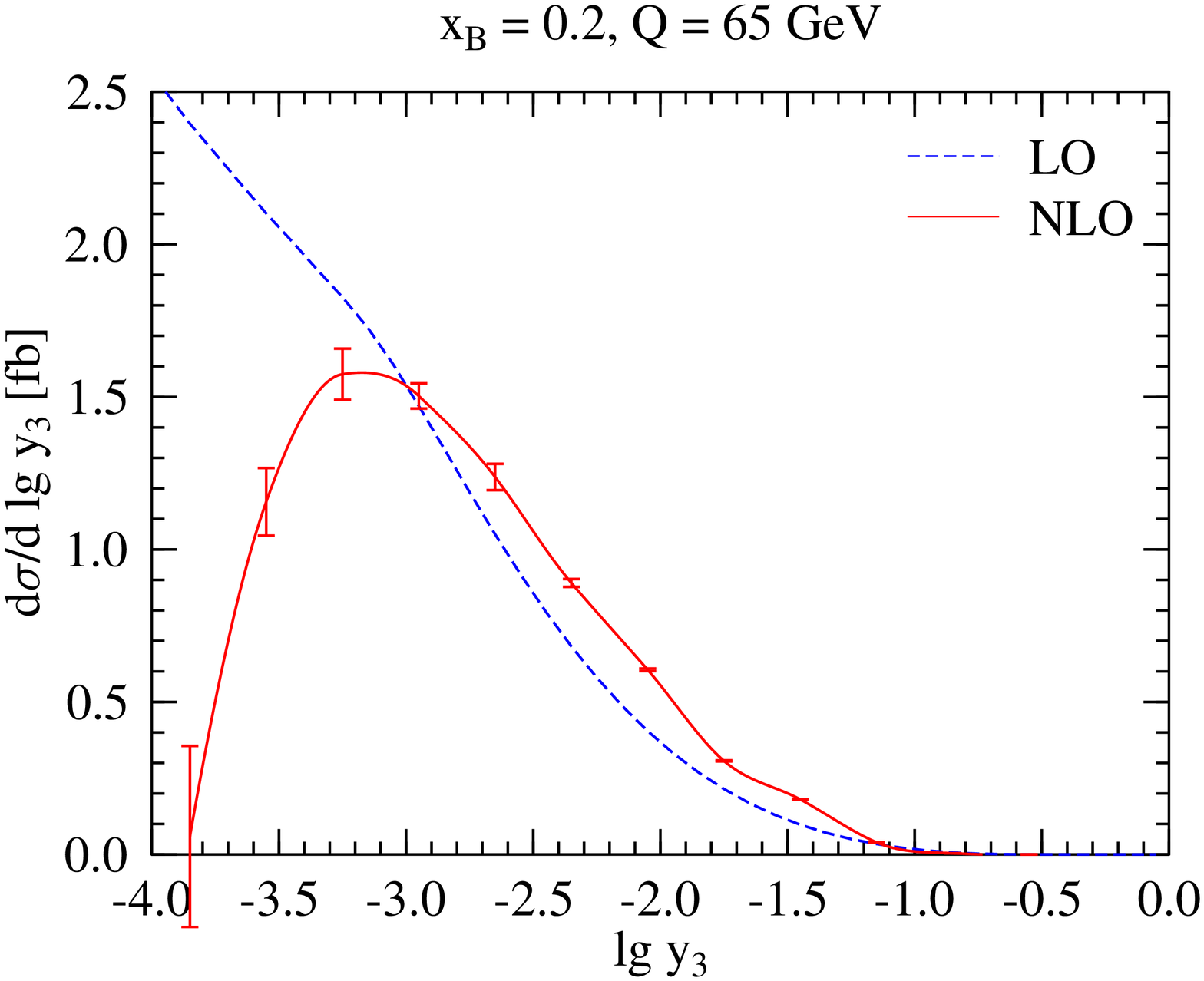}
}
\centerline{
\epsfxsize=8cm \epsfbox{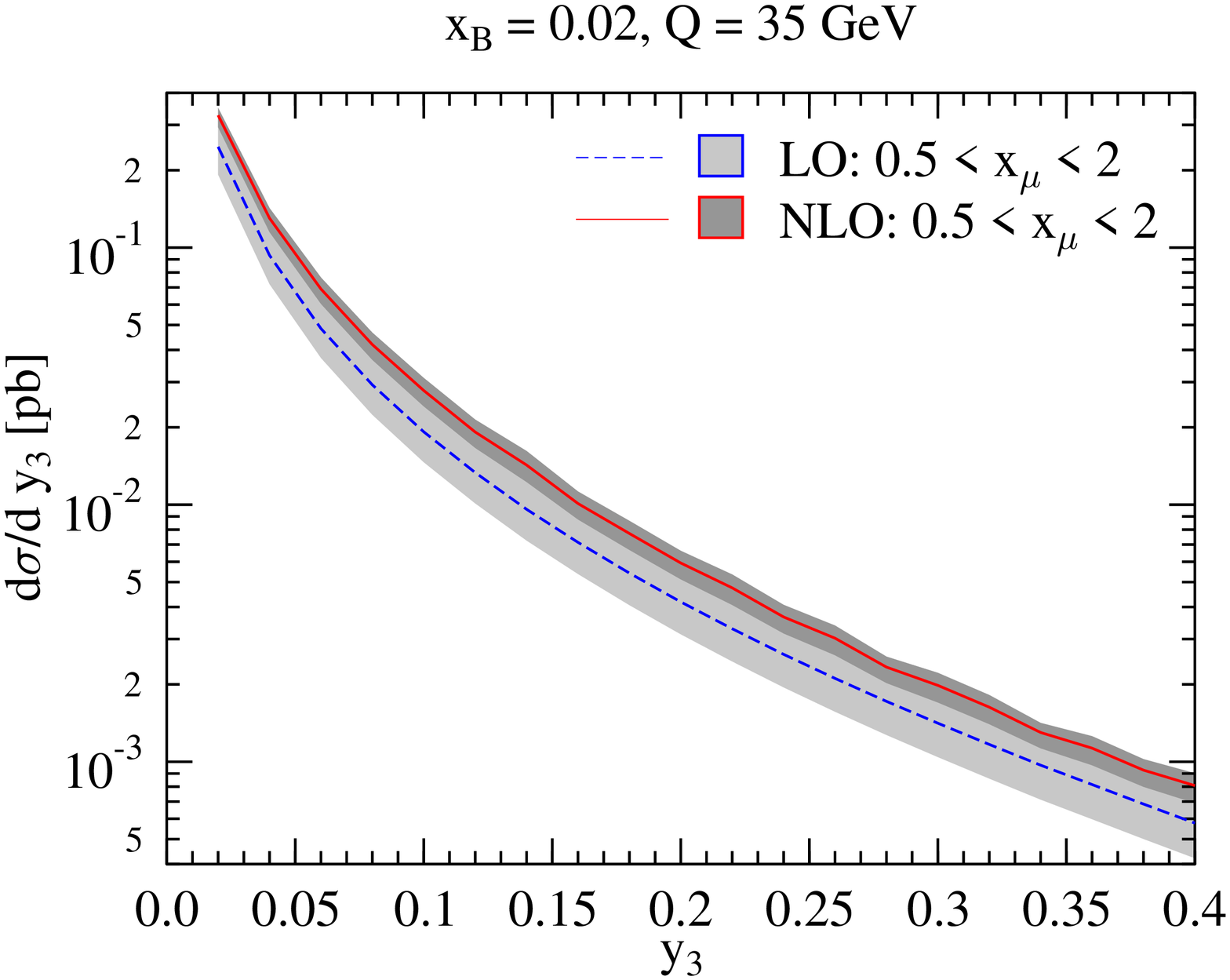}
\epsfxsize=8cm \epsfbox{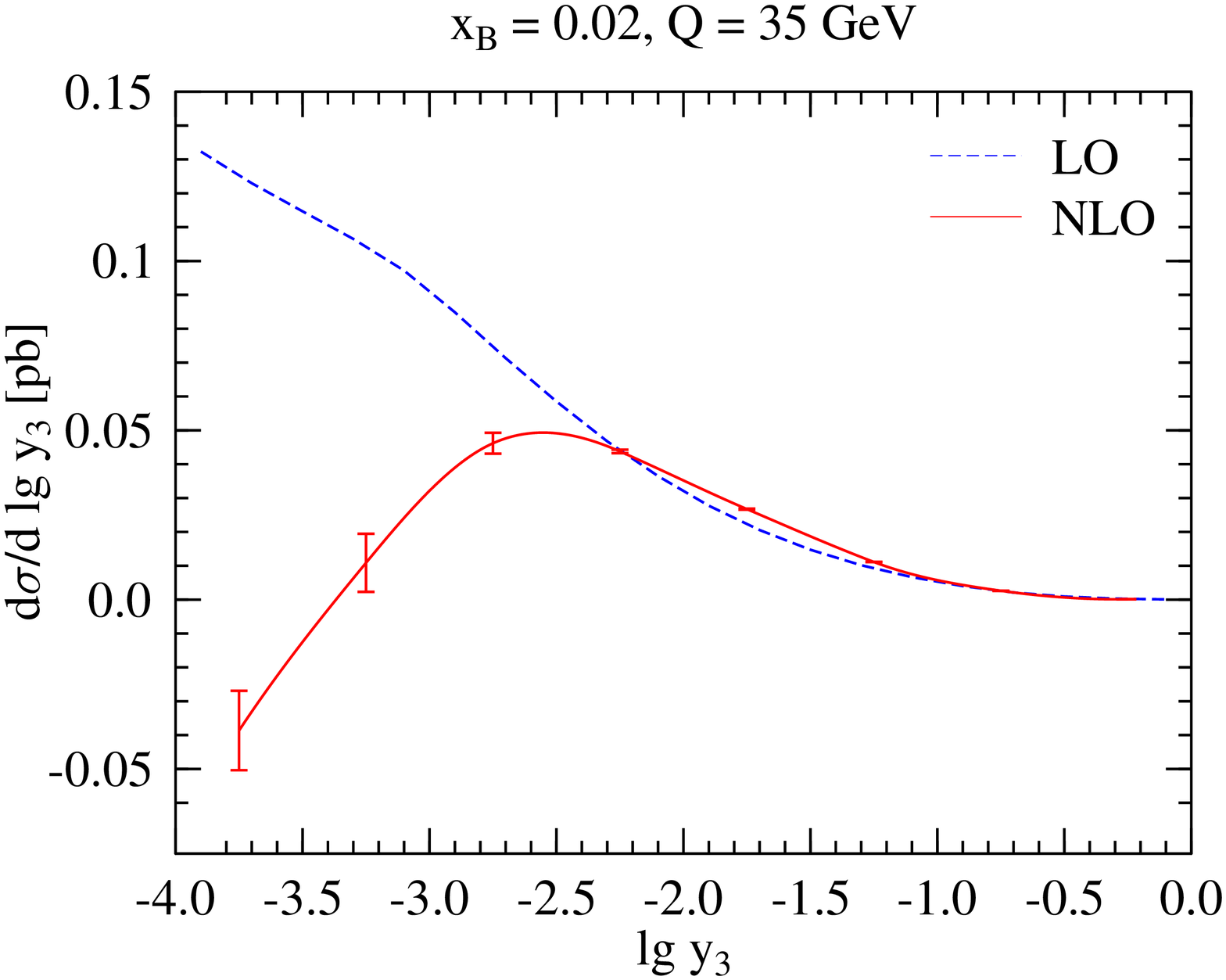}
}
\caption{The differential distribution of the $y_3$ observable at three
different fixed values of the DIS kinematic variables.  The left panel
shows the distributions as a function of $y_3$, the right panel shows
the distributions as a function of $\lg y_3 \equiv \log_{10} y_3$ in
order to exhibit the logarithmic dominance for small values of the
observable.  The \LO\ predictions are shown with dashed lines, the
\NLO\ predictions are shown with solid lines. The errorbars indicate
the uncertainty of the numerical integration that is negligible for the
computation at LO accuracy.}
\label{fig:y3}
\end{figure}

\section{Conclusions}

In this paper we presented a computation of NLO corrections to
the differential distributions of the three-jet event-shape observables 
$\Kout$ and $y_3$ in DIS. We found large radiative corrections,
especially in the case of $\Kout$, indicating that the inclusion of
even higher order corrections as well as the non-perturbative power
corrections is necessary in order to make a reliable prediction, useful
for experimental analysis. The cross sections decrease rapidly with
increasing values of the event-shape variables, leaving the region of 
small values of the observables with sufficient statistics for an
experimental analysis of data collected at HERA. In
these regions the all-order resummed predictions in the NLL
approximation are known, therefore, the matching of the NLO and NLL
distributions promises us reliable predictions. Such results are expected
to be available soon \cite{Banfiprivate}.

\medskip
We are grateful to G. Zanderighi and A. Banfi for their helpful
correspondence on DIS event-shape observables.
This work was supported by the Hungarian Scientific Research
Fund grant OTKA T-038240 and by the Swiss National Science Foundation
(SNF) under contract number 200020-109162.


\begin{thebibliography}{99}
\bibitem{DSevshape}
  M.~Dasgupta and G.~P.~Salam,
  J.\ Phys.\ G {\bf 30}, R143 (2004)
  [arXiv:hep-ph/0312283].

\bibitem{NTQCD98}
Z.~Nagy and Z.~Tr\'ocs\'anyi,
Nucl.\ Phys.\ Proc.\ Suppl.\ {\bf 74}, 44 (1999)
[hep-ph/9808364].

\bibitem{DSDIS}
  M.~Dasgupta and G.~P.~Salam,
  JHEP {\bf 0208}, 032 (2002)
  [arXiv:hep-ph/0208073].

\bibitem{HERAevshapes}
  C.~Adloff {\it et al.}  [H1 Collaboration],
  Eur.\ Phys.\ J.\ C {\bf 14}, 255 (2000)
  [Erratum-ibid.\ C {\bf 18}, 417 (2000)]
  [arXiv:hep-ex/9912052];

  S.~Chekanov {\it et al.}  [ZEUS Collaboration],
  Eur.\ Phys.\ J.\ C {\bf 27}, 531 (2003)
  [arXiv:hep-ex/0211040].

\bibitem{HERA3jets}
  C.~Adloff {\it et al.}  [H1 Collaboration],
  Phys.\ Lett.\ B {\bf 515}, 17 (2001)
  [arXiv:hep-ex/0106078];

  S.~Chekanov {\it et al.}  [ZEUS Collaboration],
  arXiv:hep-ex/0502007.

\bibitem{NTdisPRL}
  Z.~Nagy and Z.~Tr\'ocs\'anyi,
{\it Phys.\ Rev.\ Lett.}\  {\bf 87}, 082001 (2001)
  [arXiv:hep-ph/0104315];

\bibitem{Kout}
  A.~Banfi, G.~Marchesini, G.~Smye and G.~Zanderighi,
  JHEP {\bf 0111}, 066 (2001)
  [arXiv:hep-ph/0111157].

\bibitem{caesar}
  A.~Banfi, G.~P.~Salam and G.~Zanderighi,
  JHEP {\bf 0503}, 073 (2005)
  [arXiv:hep-ph/0407286].

\bibitem{caesarpage}
A.~Banfi, G.~P.~Salam and G.~Zanderighi, {\sc caesar} homepage: 
qcd-caesar.org.

\bibitem{Everett}
A. Everett, 
``Event shapes in deep inelastic $ep\to eX$ scattering at HERA'',
Proceedings of the XIII International Workshop on Deep Inelastic
Scattering.

\bibitem{kTclusDIS}
S.~Catani, Y.~L.~Dokshitzer and B.~R.~Webber,
Phys.\ Lett.\ B {\bf 285}, 291 (1992).

\bibitem{nlojet}
Z.~Nagy, {\sc nlojet++} homepage: www.cpt.dur.ac.uk/$\sim$nagyz/nlo++/.

\bibitem{CSdipole}
S.~Catani and M.~H.~Seymour,
Nucl.\ Phys.\ B {\bf 485}, 291 (1997)
[Erratum-ibid.\ B {\bf 510}, 291 (1997)]
[hep-ph/9605323].

\bibitem{NT4jet}
Z.~Nagy and Z.~Tr\'ocs\'anyi,
Phys.\ Rev.\ D {\bf 59}, 014020 (1999)
[Erratum-ibid.\ D {\bf 62}, 014020 (1999)]
[hep-ph/9806317].

\bibitem{Npp3jet}
  Z.~Nagy,
  Phys.\ Rev.\ D {\bf 68}, 094002 (2003)
  [arXiv:hep-ph/0307268].

\bibitem{Kleiss:1994qy}
  R.~Kleiss and R.~Pittau,
  Comput.\ Phys.\ Commun.\  {\bf 83}, 141 (1994)
  [arXiv:hep-ph/9405257].

\bibitem{BDK}
Z.~Bern, L.~Dixon, D.~A.~Kosower and S.~Weinzierl,
Nucl.\ Phys.\ B {\bf 489}, 3 (1997)
[hep-ph/9610370];
%
Z.~Bern, L.~Dixon and D.~A.~Kosower,
Nucl.\ Phys.\ B {\bf 513}, 3 (1998)
[hep-ph/9708239].

\bibitem{Giele:2001mr}
  W.~T.~Giele, S.~A.~Keller and D.~A.~Kosower,
  arXiv:hep-ph/0104052.

\bibitem{Pumplin:2002vw}
J.~Pumplin, D.~R.~Stump, J.~Huston, H.~L.~Lai, P.~Nadolsky and W.~K.~Tung,
  JHEP {\bf 0207}, 012 (2002)
  [arXiv:hep-ph/0201195].

\bibitem{matching}
S.~Catani, L.~Trentadue, G.~Turnock and B.~R.~Webber,
  {\it Nucl.\ Phys.}\ B {\bf 407} (1993) 3.

\bibitem{Banfiprivate}
A.~Banfi and G.~Zanderighi, private communication.
\end{thebibliography}
\end{document}